\documentclass[twocolumn]{IEEEtran}
\usepackage{setspace}
\usepackage{cite}
\ifCLASSINFOpdf
\usepackage[pdftex]{graphicx}
\else
\usepackage{graphicx}
\fi
\usepackage{lipsum} 
\usepackage{amsmath,amssymb,amsthm}
\usepackage{amsfonts}
\usepackage{algorithm,algorithmic}
\usepackage{array}
\usepackage{makecell}       
\usepackage{cite}
\usepackage{esint}
\usepackage{enumerate}
\usepackage{times}
\usepackage{url}
\usepackage{color}
\usepackage{epstopdf}
\usepackage{float} 
\usepackage{hyperref}
\ifCLASSOPTIONcompsoc
\usepackage[caption=false,font=normalsize,labelfont=sf,textfont=sf]{subfig}
\else
\usepackage[caption=false,font=footnotesize]{subfig}
\fi
\usepackage[utf8]{inputenc}

\newcommand{\PreserveBackslash}[1]{\let\temp=\\#1\let\\=\temp}
\newcolumntype{C}[1]{>{\PreserveBackslash\centering}p{#1}}
\newcolumntype{R}[1]{>{\PreserveBackslash\raggedleft}p{#1}}
\newcolumntype{L}[1]{>{\PreserveBackslash\raggedright}p{#1}}

\begin{document}
\newpage
\section*{Disclaimer and Notice of the latest version}
Thanks for reading our paper at ArXiv. The latest version of BeepTrace work is presented free to you on IEEE Xplore via  \href{https://ieeexplore.ieee.org/document/9203904}{here} or copy the link \url{https://ieeexplore.ieee.org/document/9203904}.

Please note that, due to copyright reason, we cannot provide our readers with the accepted version of BeepTrace without IEEE permission, but the work is open access to all via IEEE Xplorer.
\title{BeepTrace: Blockchain-enabled Privacy-preserving Contact Tracing for COVID-19 Pandemic and Beyond}


 	\author{Hao Xu, Lei Zhang, Oluwakayode Onireti, Yang Fang, William Bill Buchanan, Muhammad Ali Imran
	
 	\thanks{
 		H. Xu, L. Zhang (corresponding author), O. Onireti and M. A. Imran are with the James Watt School of Engineering, University of Glasgow, UK; E-mail: h.xu.2@research.gla.ac.uk;   Lei.Zhang@glasgow.ac.uk; Oluwakayode.Onireti@glasgow.ac.uk; Muhammad.Imran@glasgow.ac.uk. 
 		
	Y. Fang is with the Business School, University of Aberdeen, UK; E-mail: r01yf17@abdn.ac.uk.
	
		W. B. Buchanan is with Centre for Distributed Computing, Networks and Security at Edinburgh Napier University, UK; Email: B.Buchanan@napier.ac.uk
 	
 	}    
 }

\maketitle

\begin{abstract}
The outbreak of COVID-19 pandemic has exposed an urgent need for effective contact tracing solutions through mobile phone applications to prevent the infection from spreading further. However, due to the nature of contact tracing, public concern on privacy issues has been a bottleneck to the existing solutions, which is significantly affecting the uptake of contact tracing applications across the globe. 
In this paper, we present a blockchain-enabled privacy-preserving contact tracing scheme: BeepTrace, where we propose to adopt blockchain bridging the user/patient and the authorized solvers to desensitize the user ID and location information. Compared with recently proposed contract tracing solutions, our approach shows higher security and privacy with the additional advantages of being battery friendly and globally accessible. Results show viability in terms of the required resource at both server and mobile phone perspectives. Through breaking the privacy concerns of the public, the proposed BeepTrace solution can provide a timely framework for authorities, companies, software developers and researchers to fast develop and deploy effective digital contact tracing applications, to conquer COVID-19 pandemic soon. Meanwhile, the open initiative of BeepTrace allows worldwide collaborations, integrate existing tracing and positioning solutions with the help of blockchain technology \footnote{\textcolor{black}{Note:  This  work  has  been  submitted  to  the  IEEE  for possible  publication.  Copyright  may  be  transferred  without notice, after which this version may no longer be accessible.}}.  

\end{abstract}

\begin{IEEEkeywords}
 COVID19, Coronavirus, Digital contact tracing, Privacy-preserving, Distributed system, Blockchain, Distributed Ledger Technology, Pandemic, BeepTrace
\end{IEEEkeywords}

\IEEEpeerreviewmaketitle

\section{Introduction}







Coronavirus disease 2019 (COVID-19) is an infectious disease that is caused by severe acute respiratory syndrome coronavirus 2 (SARS-CoV-2) \cite{Gorbalenya2020}. The disease has spread into most nations across the globe thus sending billions of people into lockdown as health services across the globe struggle to cope. As of 18th May 2020, there have been 4,769,177 cases and 316,898 deaths confirmed across 188 countries and territories \cite{Worldometers}. At the time of writing this paper, there are still no vaccines for COVID-19. Hence, non-pharmaceutical interventions (NPIs), which aims at slowing down the transmission of the disease by reducing the contact rate of people in the general public \cite{Bootsma2007}, have been implemented by various countries across the globe. NPIs largely targets social distancing (also known as physical distancing) by keeping a certain distance from others and avoiding gathering together in large groups \cite{HarrisMargaret2020}. Strict measures are adopted in most countries include the closing of workplace, schools, social venues, and travel restrictions, etc. 



\begin{table*}[h]
\caption{Comparisons of current contact tracing solutions with Proposed Blockchain-based solution (BeepTrace)}
\centering
\label{tab:compare}
\resizebox{\textwidth}{!}{%
\begin{tabular}{|l|c|c|c|c|c|c|c|}
\hline
Name of solutions     &Positioning/grouping technology     & \begin{tabular}[c]{@{}c@{}}Power Usage\end{tabular} & \begin{tabular}[c]{@{}c@{}}Security of\\ technology \end{tabular}& Coverage & Privacy-preserving\\
\hline
Singapore TraceTogether\cite{Bay2020}         &Bluetooth  & High        & Low      & Low      & No                       \\
Google/Apple Contact Tracing \cite{AppleGoogle} &Bluetooth & High       & Low      & Low      & Yes, Partially                    \\
UK NHS Contact Tracing   \cite{Snow2020}        &Bluetooth & High        & Low      & Low      &  Yes, Partially             \\
China Health Code System \cite{Mozur2020}      &GPS, QR code & Low         & Medium      & High     & No                   \\
BeepTrace (proposed solution)                &GPS, Bluetooth, Cellular and WiFi   & Medium      & High     & High     & Yes                 \\
\hline
\end{tabular}%
}
\end{table*}
NPIs were found to be very effective in the H1NI influenza pandemic (1918-1919), which was the last disease pandemic at the scale of the COVID-19 pandemic, and without existing vaccines \cite{Ferguson2020}. Communities and cities that implemented NPI early in COVID-19 pandemic are successfully reducing the number of cases while the adopted measure remaining in place. This resulted in a significant reduction in the mortality rate. However, strict measures pose an immediate threat to the economy. This matters as economic decline itself has an adverse effect on many aspects of society including health. Goldman Sachs has predicted that the US economy could shrink by $24\%$ in the second quarter of 2020, more than twice as much as any decline ever recorded \cite{mckee2020if}. 
Most countries across the globe are developing balanced strategies to take both economy and a rebound of the COVID-19 into consideration. Contact tracing has been a pillar of communicable disease control in public health for decades and shows its effectiveness on COVID-19 control in some countries. With no obvious prospect of vaccines on the horizon \cite{CEPI2020}, the strategy of most governments across the globe (e.g. US, Spain, UK, Italy, Germany, etc.) for easing out the social distancing restrictions centers more on track and trace approach. This approach will further help to rescue the economy while also saving lives and restoring some normality, especially when the lockdown is lifted (or partially lifted) and the society steps into a ``New Normal".

\subsection{Contact tracing}\label{contact-tracing-f}


Contact tracing is the process of identifying  persons who may have come into contact with an infected person and subsequent collection of further information about these contacts \cite{Ferrettieabb6936}. Contact tracing has a long history in preventing infectious diseases, in the early stage of epidemiology, contact tracing takes part with labor-intensive methods. The process relied heavily on the recall of a (far from complete) list of people whom they have been in contact with over the previous weeks, or locations the confirmed person has been. Letters, phone calls or emails can be used to inform people who might be contacted. Thus, completeness and accuracy of the list, timeliness and efficiency of the tracing are limited by such a traditional contact tracing approach.

Until very recently, digitized contact tracing through smartphone apps are developed and deployed in some countries to solve the bottlenecks of the labor intensive-methods. One of the mainstream contact tracing approaches is to use Bluetooth signals from smartphones to detect encounters with people with COVID-19. 
This approach does not use location tracking or store users' location data. 
In this approach, if someone develops COVID-19 symptoms, an alert could be sent to others that they might have infected, with minimum intervention. There are two variants of the Bluetooth-based contract tracking, namely, the centralized and the decentralized model. 
Singapore’s TraceTogether\cite{Bay2020} is an example of a centralized model. On the other hand, the information is kept on the user’s smartphone in the decentralized model and this gives more control to the user. 
Processing and matching for people who may have contacted COVID-19 are made on the user’s smartphone in a decentralized model. Moreover, the decentralized model has been promoted by an international consortium including Google and Apple as it promotes consent, transparency, and privacy \cite{AppleGoogle}. In the former, gathered anonymous data is uploaded to the server. Matches are made with other contacts via processing on the server if someone starts to develop the COVID-19 symptoms. For simplicity, in the next, we omit the word ``digital'' and use the term ``contact tracing'' representing the smartphone App-based digital contact tracing. 

\subsection{Review of existing contact tracing solutions}\label{sec:compare}
In Table \ref{tab:compare}, we review four of the most recently proposed contact tracing approaches namely, TraceTogether from Singapore\cite{Bay2020}, Google/Apple joint contact tracing project\cite{AppleGoogle}, NHS COVID-19 App\cite{Snow2020}, and China Health code system\cite{Mozur2020}. The metrics used in our evaluation include the positioning or grouping technology, power usage, security of the technology, coverage, and the level of privacy preservation.

TraceTogether is an App powered by Bluetrace\cite{Bay2020} protocol and it makes use of Bluetooth low energy (LE) to discover and locally record clients in close proximity of a user. In this scheme, the user is required to keep the device in an active broadcasting state, hence drains the battery of the user device. The Bluetooth technology has security concerns on its vulnerable wireless interface, threats including bugging, sniffing, and jamming are prominent to all Bluetooth-based contact tracing solutions. There is a high risk of replay attacks to the contact tracing network, which may later cause a massive scale of panic to the public.

 Meanwhile, the Bluetooth protocol is potentially liable to be used against user security, since the identification of hardware on the Bluetooth physical layer may not be concealed, which brings the exposure of the physical hardware. 

Such privacy may be considered preserved in terms of the macro-scale public, but it is almost transparent inside a local group with Radio Frequency vulnerability. Meanwhile, the problem of locally initiated proximity solutions is limited due to the transmission power limit on the user device and existing wireless interference. 
As we discussed earlier, TraceTogether is a centralized service in terms of the user's real identity and notification, though user privacy is not known to the third party but the authority. Therefore, it is considered not genuinely privacy-serving, if the malicious activity is by the central service provider.

Google Apple Contact Tracing employs a similar approach with Bluetooth LE too. It is different from TraceTogether from the user's privacy perspective since the service provider does not get hold of the user's real identity, hence becomes privacy-preserving. However, the user is required to use their central server for contact matching and notification, which brings the concern of trajectory attack on user privacy and enables the reconstruction of the user's profile using access information to the server. Similarly, the NHS COVID-19 App has risks of potential exposures of user privacy in the same way. 

Health Code System is different from the above methods, as it does not use Bluetooth nor proximity detection. It is based on relational cross-match by scanning the QR code, which is associated with the user. In this system, user privacy is not respected due to centralization, and the identity of the user is not hidden to the authority. However, the health code is only scanned at the time of passing checkpoints, hence saves the user battery and does not consume data. Additional, thanks to its highly central hierarchy, the coverage can be extended easily. 

Many other protocols and solutions are emerging to deal with pandemic contact tracing, such as \textit{Aarogya Setu}, \textit{COVIDSafe}, \textit{Decentralised Privacy-Preserving Proximity Tracing}, \textit{Pan-European Privacy-Preserving Proximity Tracing}\cite{IndiaGoverment2020,DepartmentofHealthAustralia2020,DP-3T2020,PePP-PTe.V.i.Gr2020}, etc. They are similar to the solutions described above with their tweaks on certain features.

	\subsection{Blockchain basis for contact tracing}\label{blockchain-platform-selectioncomparison-p-fig.1-blockchain-consensus-from-ks-work}
The nature of contact tracing brings challenges in privacy since the information has to be collected, matched and distributed. 
Other issues include guaranteeing the protection of the identity of users with COVID-19. Though the opt-in option could ensure some control over participation, it is yet to be seen how we ensure that only the relevant data are shared. Blockchain can play a neutral role in a distributed manner to bridge the user/patient and the authorized solvers to desensitize the user ID and location information. It can provide a solution for privacy-preserving from technical design rather than relying on the obey the regulations or laws in a centralized system.  
 Furthermore, blockchain technology when combined with the use of encryption and anonymization technologies can further protect the users' identity. Blockchain in nature is non-regional thus provide a suitable 
global access platform for COVID-19 pandemic tracing and control. The transparency feature can prevent the public from intentional misinformation by authorities or other third parties. 
More details about blockchain will be provided in Section \ref{sec:blockchain}.


	\subsection{Motivation}\label{motivation}
Recognizing the challenges and issues above, enhanced privacy preservation, better tracing performance, and better capability to fight against misinformation are required for the post-pandemic contact tracing. We deem that no compromise should be made between the privacy and tracing performance, hence we present our blockchain-enabled contact tracing that satisfies both privacy and performance requirements. We have summarized these aspects below:
\paragraph{Enhanced privacy as the main focus}
Contact tracing in nature is sensitive to the general public's privacy and security, hence the privacy should be respected in the solution framework design. It is the most concerning factor in all contact tracing proposals we have seen recently. Meanwhile, and the more information collected, the better performance of contact tracing. However,  privacy should never be sacrificed.  
Meanwhile, the secured data sharing is another challenge for privacy, it would be a hard decision to make, choosing between health and the consent of centralized privacy collection, since the centralization brings the risk of manipulation and corruption. Nevertheless, it is not a trouble for blockchain, where the identity is removed at the beginning, offering the tracing participants with ultimate confidence in privacy.

\paragraph{No compromise of tracing performance}
By preserving the privacy from the users, we believe the performance of contact tracing also matters. The performance of the tracing network should be valued from its effectiveness of infection prevention, including the level of technology and the coverage of the network. The current decentralized solutions are limited to a local network, hence do not have impacts on a wider range of users. For example, people who travel across different areas for work and leisure in the post lockdown period may benefit from a wider range of tracing. Blockchain in nature can be the key to enable the globally accessible tracing network. It is also a challenge for government-initiated tracing projects or Apple Google joint effort\cite{AppleGoogle,Bay2020,Snow2020}, since their reach is limited either by political or technical reasons, for instance, Google has no accessibility in China. We aim to utilize the blockchain, making all users connected to the chain without violating their privacy. Moreover, to make use of all the possible tracing information, a framework for supporting all means of positioning technologies is required. The information shared on the blockchain can hence be propagated further and is a lot richer than Bluetooth interactions. 

\paragraph{No panic from misinformation}
Misinformation is harmful to pandemic prevention and causes panic to the general public. The main reason for the misinformation can be concluded into two categories: information inaccuracy and information transparency. 
The public health agencies have strong reasons to get the trusted authorities involved in the result confirmation, geographical matching, and notification, in order to fight against misinformation of inaccuracy. 
Though getting the authorities involved might not sound promising for the fact of privacy, it is not the case for the blockchain network, thanks to the privacy-preserving ability. The panic should not be caused by the tracing and never will be. 
Meanwhile, the authorities have the motivation to hold back information or provide false statistics due to their favor of decisions enabled by the centralization of data. With the help of the transparency provided by the blockchain technology, it enables the easy verifiable trusted tracing information by the public rather than a closed group of informants. 

	\paragraph{Full life cycle privacy protection}
It is acknowledged widely that privacy should be valued from the start to its end, hence a full life cycle solution of privacy for contact tracing is necessary. The shared data should have its life cycle managed from users' tip of fingers. The users of contact tracing shall have full privilege to share and revoke sharing at any time using key management. Public agencies are also required to limit the sharing of users' sensitive information within a trusted partnership. The proposed blockchain platform is capable of providing users and agencies with thorough credential management functionality with cryptography. The user's privacy is protected throughout the tracing scheme, and the length of data storage should also be regulated under GDPR (General Data Protection Regulation)\cite{GDPR}, and health agencies' recommendations, for example, WHO recommended that 14 days is the minimum length of tracing cycle. In short, the user will have all its privacy protected through generation, sharing, and deposition.

\begin{figure*}
	\includegraphics[width=18cm]{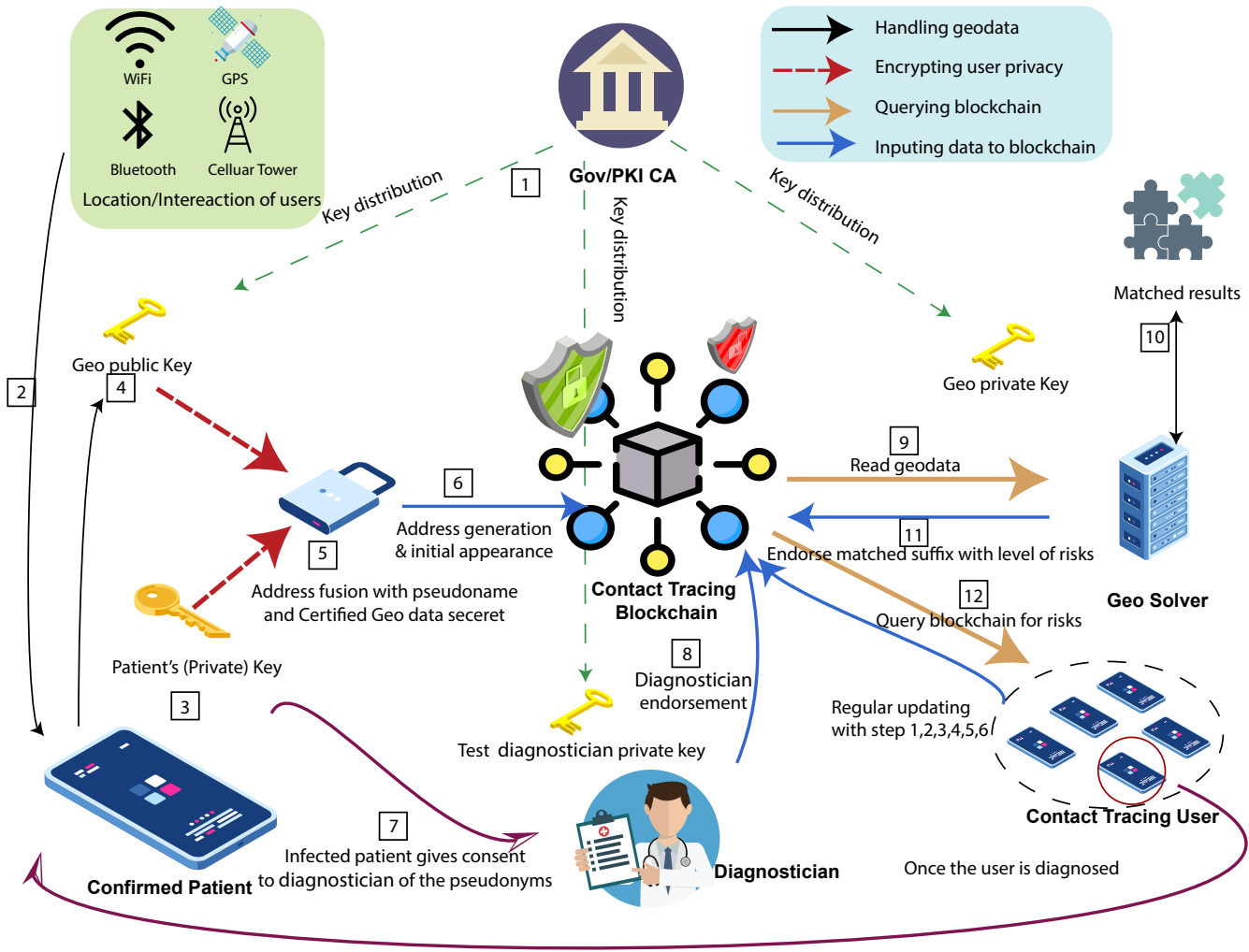}
	\caption{Framework of Blockchain-enabled Privacy-preserving contact tracing scheme (BeepTrace)}
	\label{fig:frame}
\end{figure*}

	\subsection{Contributions}\label{sec:contribution}
Considering all the motivations listed above, there is a need for a novel solution that copes with issues in existing labor-intensive and restrained Bluetooth contact tracing solutions.
This paper proposes a well defined Blockchain-EnablEd Privacy-preserving contact Tracing (BeepTrace) for
maximum privacy preservation, making an efficient network of contact tracing, breaking the information
barriers without scarifying privacy. User privacy is respected with full life cycle awareness, as it can be generated, shared, and disposed of safely. Meanwhile, BeepTrace proposes an architectural view of blockchain address and transaction design making use of two chains, to facilities the analysis of geodata and passive notification. 
We propose a novel scheme to decouple user privacy by using two distributed blockchains. The tracing chain with desensitized personal location information is accessible by authorized solvers for contact matching. And the notification chain, where the match results (only pseudonym or its fingerprint) will be published on for the exposed users-self matching locally. Through this Gemini chain design, all users' privacy can be preserved effectively. 
We also provide numerical results to give an overview of the network storage and computing capacity requirement with typical parameters setting. 
In addition to network cost, an analysis of data consumption for an individual user is also calculated to address the concerns of the device requirement.

The solution proposed in Fig. \ref{fig:frame} of this paper provides an open initiative framework for governments, authorities, companies, software developers and researchers around the world to develop and deploy a fast and trusted platform for tracing information sharing, to minimizes the damage COVID-19 does to humanity and to save lives and economy without invading the basic human rights of privacy.


\section{Blockchain as the backbone for privacy-preserving information sharing} \label{sec:blockchain}

Blockchain technology, which has shown great potentials in various fields such as financial services, energy trading, supply chain, identity management, and the Internet of Things (IoT) \cite{Cao2019,Xu2020} could address the trust, privacy, security, and transparency issues associated with the existing contact tracing technologies. Blockchains are distributed databases organized using a hash tree, which is naturally tamper-proof and irreversible \cite{Underwood2016}. In particular, data introduced into the blockchain platform are organized into blocks. Each block has an associated hash value for that block, this applies to the previous block as well and thus ensures a retroactive linkage between blocks. Blockchain offers an immutable, transparent, secure, and auditable ledger in a trust-less distributed environment, to verify the integrity and tractability of information/assets during their life cycle.

Blockchain can be integrated into contract tracing applications to provide much need security, trust, transparency, and privacy which are either missing or partially provisioned in the existing schemes. Besides its chain-link data structure nature, the Consensus Mechanism (CM) is of great importance to achieving the unique gains of blockchain. The CM ensures an unambiguous ordering of transactions and guarantees the integrity and consistency of the blockchain across geographically distributed nodes. The CM largely determines blockchain system performance, such as transaction throughput, delay, node scalability, and security level, etc. As such, depending on application scenarios and performance requirements, different CMs have been considered for blockchain. Important requirements to be considered when selecting the CM in BeepTrace include network throughput, delay, storage, and scalability. Commonly used CMs include Proof of Work (PoW), Proof of Stake (PoS) and Direct Acyclic Graph (DAG) based CM. 

 PoW was proposed in the original blockchain application (Bitcoin) and its core idea is the competition of computing power. Each node involved in the CM uses its computing resource for the hash process to compete for the right to the new block while receiving some bonus. This leads to the use of computing resources and meaningless energy consumption. 
PoS, on the other hand, relies on coin age competition rather than computing power competition. PoS is thus beneficial for the wealthy miner and it could cause near-monopolies, which can result in the generation of a powerful third-party. This could also be a challenge in BeepTrace where the users’ privacy at stake. By design, a more balanced weighting scheme on coin age can solve such a problem. Both PoW and PoS CM work on a single chain architecture. To maintain a single version of the blockchain among the users, the CM must reduce the access rate of new blocks \cite{Cao2019}. This could lead to some bottlenecks in applying PoW and PoS CMs to a large number of contact tracing participants (e.g., a country with a large population like China, India).

In particular, to reduce the access rate of new blocks and prevent the PoW or PoS based BeepTrace system from attack, the CM will consume many resources, which is too costly for such a resource-constrained system. Furthermore, with the limited capacity of the new blocks in PoS and PoW, the system will be unable to cope with the exponential growth in the number of users. For instance, the throughput  is normally limited to 7 transactions per second (TPS) in Bitcoin and 20 to 30 TPS in Ethereum \cite{ Bendiksen2018}. The low access rate of new blocks in PoW and PoS CMs implies a long confirmation delay for the CMs. Typical confirmation delays of 60 minutes in Bitcoin and three minutes in Ethereum are too long for the BeepTrace system since other delays within the network such as access delay, processing delay must be incorporated as well. Nevertheless, the throughput and delay performance can be significantly enhanced by reducing the difficulty level of harsh calculation security level (e.g., in BeepTrace, there is no need to wait for the block confirmation after 6 blocks are generated after it). Furthermore, a small to medium size population city could be used for contact tracing.

DAG-based CM can overcome the shortcomings of PoW and PoS consensus when applied in BeepTrace. Unlike PoW and PoS, there are no competitions to create a new block in DAG and all transactions are connected directly or indirectly. DAG-based consensus mechanism allows users to insert their blocks into the blockchain at any time, as long as they process the earlier transactions. In this way, many branches would be generated simultaneously, which is referred to as forking. With forking the confirmation rate and the TPS are both unlimited in DAG-based CM. Moreover, with the forking integrated into DAG, the resource consumption can be very low for a user to create a new block, thus making it very suitable for the BeepTrace system. Other key benefits of DAG-based CM which make it more suitable for BeepTrace include zero
transaction fees and low computing power \cite{Cao2019}.

\section{Blockchain EnablEd Privacy-Preserving Contact Tracing }\label{sec:framework}
In this section, we give a detailed workflow description and an explanation of key concepts. In the following parts, we will first introduce the entities involved in the system, their roles, and how they interface with each other.   
The workflow of the contact tracing framework will be proposed next, then we will explore the details of blockchain pseudonym generation and sharing. 

It is worth noting that the framework works with an open initiative that allows everyone to share the contact tracing information with different methods, authorities, and cryptography, and it can become a piece of open interface information tracing hub for all privacy-preserving contact tracing providers globally. Moreover, the proposed framework does not limit the selection of blockchain CM and the incentive mechanism of the blockchain. As long as the CM fits the network's performance requirement, it can be plugged into the framework. Besides, the framework does not limit the selection of positioning services.

\subsection{Entities, functions and interfaces}
In the following, we define the parties involved in BeepTrace and explain their roles and interfaces one by one: 
\begin{itemize}
    \item Users (see Fig. \ref{fig:frame}, includes confirmed patients and healthy users), is an abstract term of contact tracing App users on a mobile device. We use ``user'' to represent user equipment (UE), the App, and the device in the rest of the paper. All users will upload their encrypted  TraceCode to tracing blockchain and read from notification blockchain for self matching.  
    \item Diagnosticians (see Fig. \ref{fig:frame}), 
    diagnose and endorse confirmed COVID-19 user's geodata with a signed prefix and send to tracing blockchain for solver matching. 
    \item Geodata solvers (see Fig. \ref{fig:frame}), server or server clusters associated with the trusted third party or user, interacts with the geodata and provides endorsement on the notification chain. Reads raw data from tracing blockchain for matching and send matched data to notification blockchain. 
    \item Public Key Infrastructure (PKI)/Certified Authority (CA) (see Fig. \ref{fig:frame}),  a trusted third party (e.g., governments, public health agencies), interacts in key distribution to the user, diagnostician, and solvers.
    \item Positioning service providers (see Fig. \ref{fig:frame}), including but not limited to GPS, Bluetooth, Cellular Tower and WiFi whichever is supported by the user. Data supplied by the provider will be labeled as geodata throughout this paper.
    \item Tracing blockchain (see Fig. \ref{fig:bcaddress}), is one of two chains (will be introduced in detail in Section \ref{sec:pseudonym}) that accept TraceCode registration by user and diagnostician. It is also read by the solver for geodata matching.
    \item Notification blockchain (see Fig. \ref{fig:bcaddress}), is the chain dedicated to risk registration to the affected users' TraceCode.
    \item  TraceCode (see Fig. \ref{fig:bcaddress}), is a mask name for the blockchain address introduced in the paper, it has two parts, the front part is the user pseudonym, called prefix, the rear part is geodata cyphertext, called suffix.
\end{itemize}

\subsection{Workflow of the BeepTrace}
We explain BeepTrace using Fig. \ref{fig:frame} and we give details step-by-step.

The first step (step 1) of our proposed BeepTrace is that PKI/CA distributes the keys to the above parties, as suggested in Fig. \ref{fig:frame}.
Users will collect raw geodata from the positioning service providers, indicated with Step 2, and generate multiple local private keys over time (e.g., one for each day), these keys will be stored in users local storage, preferably, in an encrypted chip like Apple T2 Security Chip\cite{Apple2018}, as in Step 3. Such encryption will be strong enough to protect users' privacy from any known threats and avoid human mistakes. These keys will be used to generate a pseudonym, which is used as the prefix of a blockchain address, the front part of  TraceCode. Note that, both symmetric-key and asymmetric, aka. public-key encryption can be used for the user's key generation and management.

On the other hand, the user generates another cyphertext using a public key, which is certified by a CA (a trusted party), to encrypt its current geographical or topological location data with a timestamp in step 4, and forms the rear part of  TraceCode. Note that CA is introduced to provide confidence to the public but not tempering the independence of BeepTrace as it does not obtain any privacy from the user. We call this geodata cyphertext, and it will be used as a suffix of a blockchain address which is associated with the pseudonym, stated as address fusion in Step 5. 
\begin{figure*}
\includegraphics[width=18cm]{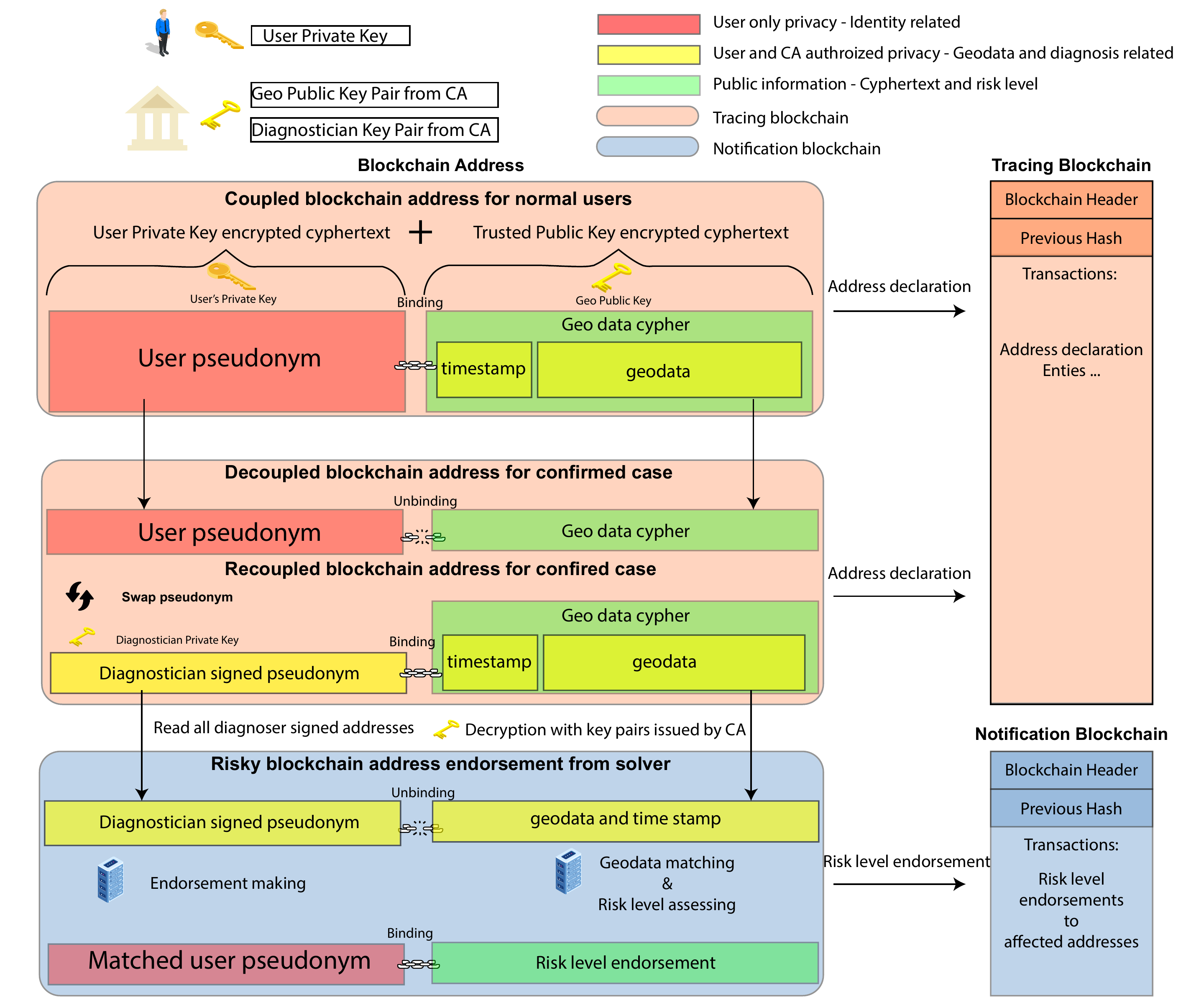}
\caption{BeepTrace blockchain address architecture (TraceCode)} 
\label{fig:bcaddress}
\end{figure*}
At this point, we have successfully established the first link of a user pseudo-identity and the geodata in the form of blockchain addresses. Once the address is generated, the user will declare it on the blockchain network (see details in Section \ref{sec:pseudonym}) in Step 6, hence the address becomes index-able using its suffix by the trusted third party, and the users' privacy remains protected due to the anonymous identity by the pseudonym. Note that, all users in the network will repeat steps 1 to 6 until the user is diagnosed with COVID-19. The following steps are for the confirmed patient.

Once a user is diagnosed by a diagnostician, the user has options to exchange its existing pseudonyms with the current handler by giving the patient's consent to this very specific diagnostician in step 7. After receiving all the pseudonyms from the users, this diagnostician tracks down all the related addresses using the prefix thanks to the users. During the pseudonym exchange, the trusted one needs to verify the pseudonyms (see details in Section \ref{sec:pseudonym}). Meanwhile, this trusted person decouples all the user private key related prefix from the geodata suffix, and replace the pseudonym section with another private key encrypted cyphertext, which is designated to the diagnosticians. The diagnostician generates a new blockchain address by re-coupling the new prefix and new suffix, which can be generated with man-made drifting/noise encryption technique for further protection, then endorse it on the blockchain network, as in step 8. 

From this point, the users' privacy is completely protected/preserved. 
The privacy is only revealed in the process of been diagnosed due to the nature of the diagnosis, and it is protected under regulations and laws, for instance, the confidentiality of the UK NHS (National Health Service) code of practice \cite{DepartmentofHealth2003} and GDPR.

After the confirmed patient's status has been updated on the tracing blockchain, illustrated in blue in Fig. \ref{fig:bcaddress}, anybody with access to the chain will be able to read the cyphertext and know the update made by the diagnostician, though access to the geodata is exclusive to the geo private key holder, issued by a public trusted parties with the previously mentioned public keys. Again, the information has no user info, at step 9 in Fig. \ref{fig:frame}. At the same time, the link of the pseudonym is only known to the user itself. 

Now we have all the required information for contact tracing (that is, an irreversible link of pseudonyms and geodata, and the diagnostician's endorsement to the geodata), the only thing left is to match them, as shown in Step 10. Any interested parties/users (in Fig. \ref{fig:frame}) who are authorized by the CA can start backtracking the geodata from confirmed patients which are marked by the diagnosticians by decryption of the geodata and  timestamp. By doing so, if cross-infection is likely between several blockchain addresses, the solvers will make an update of risk level to related addresses by looking up the suffix and endorse it on the blockchain, but they will not be able to know the user's information due to the decoupled data and pseudonyms. The marked addresses are declared again on the notification blockchain (see details in Section \ref{sec:pseudonym}) in Step 11.

As the user is using the tracing App, when the download of notification has finished locally in Step 12 (details are given in Section \ref{sec:pseudonym}), the user can look up its addresses from the notification blockchain, which is a separate chain exclusive for risk level notification, and now the users are been notified passively once the match of addresses has occurred with endorsements made to any of users' addresses. In the case of compressed results on the notification chain, the user needs to match its prefix's fingerprint with them. In this process, the user's privacy is well preserved locally, as no one without knowing the users' keys can link the user to the geodata.

A sub addresses scheme can be introduced to power the self-marking with symptoms code, without the involvement of CA. The code can be a plaintext hash and installed as the prefix of blockchain address with public key encrypted geodata. The solver can also dedicate to search the symptoms and warn others using the same technique in the previous scenario (steps 1, 2, 4, 5, 6 and 3 is dismissed due to plaintext), but the information propagated through unsigned address are not trusted, and should only be taken seriously if the community has a wide range trust basis. No personal keys are revealed in this process, hence the privacy is well preserved too.

\subsection{Blockchain pseudo-identity sharing}\label{sec:pseudonym}
In Fig. \ref{fig:bcaddress}, we illustrate the generation of blockchain addresses and the mechanism to decouple the users' privacy with the diagnosis (a signature by diagnostician) and geodata sharing. Sharing of the pseudonym generated by users' private keys are considered safe to be public. A handful of cryptography algorithms can be applied to pseudonym generation, including both symmetrical and asymmetrical encryption. In the figure, we can see that the address is divided into two parts, pseudonym prefix and geodata suffix. The user uses the private key to generate a cyphertext as its pseudonym for the front part of the address, as in Fig. \ref{fig:frame} step 3, and uses a public key offered by regional/global CA to encrypt its geodata, in step 4. It then puts the cyphertext of geodata into later part of the blockchain address as the suffix. A complete address shall provide a direct link between the pseudo-identity and the geodata in step 5. Besides, the diagnostician will need to verify whether the user is the rightful holder of pseudonyms by verifying the private keys held by the user. This ensures that the diagnosis information is shared responsibly, and is thus a critical step to avoiding public panic.

By sharing this address with the blockchain network, the information carried by the address itself will be known as a cyphertext, and potential readers will know how to separate the cyphertext into pseudonym and geodata cyphertext. However, only the authorized users/servers who have the private key from the authority can decrypt the geodata but they have no clue of the pseudonym, therefore protecting users' privacy. Interactions between tracing blockchain and notification blockchain in Fig. \ref{fig:bcaddress} are designed to offload the needs of a heavy tracing chain and enable a trusted blockchain with trusted users/servers, as only selective information will be published to this dedicated chain from trusted sources.
Meanwhile, the users' traces of internet connectivity are also concealed by blockchain, hence we can assure that what on the blockchain is nothing but a pseudonym and a geodata. Miners of the blockchain may receive connectivity trace from the user, but it is not inherited on the blockchain, either received constantly by one miner due to rapid changes of non-geographical related miners. 
In other words, users access information such as IP addresses, routing information, and even the ISP records are completely isolated from the blockchain network, hence the network is born to be real privacy-preserving. This advantage may be overkill for some countries' regulations, but it will be a gem of privacy-preserving.

\section{Geodata generalization for privacy protection and solving}\label{sec:geo}
Contact tracing blockchain is not limited to any specific geographical or topological information collected by GPS, Wifi, Bluetooth, base station, and any other indoor or outdoor positioning technologies. It is a platform that fuses all types of geodata and shares them for geodata matching. With the help of BeepTrace, the user privacy is in safe hands, however, it still faces a challenge of geodata overhearing issued. This is a critical challenge during the geodata capturing and the protection of privacy and granularity of data accuracy in addition to the secured mainframe design. We present our geodata generalization plan to guide geodata capturing and storing. 

The raw geodata is generated by GPS/Wifi/Cellular tower when the services are available to the users. Next, the users' device will upload them to the blockchain network with a public key issued by CA.
Meanwhile, the geodata will undergo perturbations here to avoid identical suffix match tracing against the patient's private key, by either adding salt to geodata or transform the geodata.
The relevant transformations can be achieved in three ways: 
\begin{itemize}
    \item Use geo datum with an elliptical encrypted system with perturbation, a well-known example of the implementation is GCJ-02 datum reference system.
    \item  Convert GPS geo datum (WGS84)\cite{DepartmentOfDefense2008} to the Grid reference system (OSNG: OSGB36, where the accuracy is limited). 
    \item GIS aggregation/ geodata generalization and perturbation to avoid trajectory privacy tracking.
\end{itemize}
The user has the freedom to choose which level of detail it intends to provide to the blockchain tracing network, as long as the accuracy level fits within the regulation made by the local agency. In addition to that, the diagnostician can use the above methods again to convert the users' geodata into a coarser grain to avoid trajectory tracing by malicious users or even completely reconstructing the geodata with a dedicated key for secured geodata sharing if required. The management of geodata allows fine-grained access control to be achieved on the blockchain.

\subsection{Geodata solving and reverse topological cross-infection warning} 
Contact tracing is closely related to geographical intersections of the traced target, which is represented as a set of geodata. A first-party or third-party solver is needed to decrypt the geodata from the cyphertext first, then run the patients' every record against the whole data collected within 14 days for COVID-19 tracing (recommended by WHO and it could be different for other pandemics). The most simple way to this is to calculate a distance-vector which will be used as the metric along with the contact duration for the risk level assessment. Besides the geographical information, the Bluetooth group information can also be integrated into the network, if the users are willing to link them. This will solve the limited proximity issues for all Bluetooth technology-based contact tracing solutions\cite{Snow2020,Bay2020} by extending the tracing to more possible positioning services, and enriching the geo solving model in the solver side.

Third-party GIS (Geographical Information System) services can be integrated. For example, any solver can link the geodata with OSM (Open Street Map)\cite{OpenStreetMap} to get GIS data from OSM and make use of the information like object type, road topology, building name and function, the height of the object, etc., which can be fused into risk level management. For instance, the road topology can indicate the trajectory, determine if the user is outdoor or indoor. Such information fusion and processing can bring the contact tracing not only beyond the geodata, but also explores social connections. Speaking of which, it is unimaginable if the privacy is exposed or hijacked by any malicious party, therefore, privacy-preserving is not an option but a must. By adding the GIS into the solving loop, the external topology can be taken into account, for example, if the geodata is found out to be at the center of a shopping mall, then the shopping mall's topological information can be obtained from OSM and used to warn the rest of the people who were in the shopping mall at the same time. This method provides more flexibility compared to solely coordinates/proximity, as portrayed in TraceTogether\cite{Bay2020} and NHSX contact tracing\cite{Snow2020}.

\subsection{Risk level management and notification} 
Once the geodata solver extracts the high profile geodata using a clinical endorsement, the matching is conducted.
The risk level, therefore, can be worked out using the government guideline on the distance and contact time. For instance, any users who were within the proximity of 10 meters more than 15 minutes will be marked as High-risk exposure\cite{EuropeanCentreforDiseasePreventionandControl2020}, and those who were further and stayed less than 15 minutes will receive a low-risk exposure endorsement. With the enhanced topological matching, the details can be set up by the solver itself with certified guides by authority, for example, if the authority thinks the indoor activity brings the risks to every people in the facility, then the topological information can be used for marking. 
 The result will be linked to the address who has an endorsement from the solver. The address is considered to be notified passively at this stage. It is worth noting that the risk level endorsement is public information, but the only way to make use of them is to look up the prefix and identify users themselves actively. The process is like a radio broadcast, and the receivers are listening to it passively. There are drawbacks of passive notifications regarding its performance (discussed in Section \ref{sec:elders}), but this will again double assure the users' privacy is protected by design (no trust is needed).
\subsection{Complete freedom of pseudonym revoking and sifting} 
In the geodata matching process, there is a risk of trajectory tracing against users' pseudonym, however, users have complete freedom to change their private key more frequently to avoid any possible leak of privacy, with the cost of increasing data consumption and storage. 

Users can renounce the private key at any time and start using a new private key at will, in order to prevent anyone from retrospectively using the logs to reveal the users' activity pattern.

In the case that a user wants to revoke shared data from BeepTrace, it can do so by informing the PKI/CA to revoke the public key assigned to its geodata, hence revoking the information back and forth.

\section{Results}\label{sec:results}
In the results, we first present an analysis of existing solutions to contact tracing, which we compare with our proposed solution referred to as BeepTrace. Then we provide numerical results to show BeepTrace's performance in terms of blockchain requirements of throughput and storage. Next, we analyze the computing resource requirement for geodata matching with some illustrative figures. In the end, we work out the user side requirement regarding data consumption and storage. 
\subsection{Comparisons with existing contact tracing solutions}
By comparing the efforts made by different countries and agencies, there are clearly two divisions both technical-wise and privacy-wise. Technical-wise, there are two types of tracing mechanisms, one uses health code while the other utilizes Bluetooth. 
Detailed comparisons of the contact tracing solutions can be seen in Table \ref{tab:compare}, where we list our BeepTrace with four widely acknowledged solutions. As discussed earlier in Section \ref{sec:compare}, Bluetooth based solutions, are energy starving for users, since the device must be kept active and broadcasting all the time to achieve such functionality.
On the other hand, the health code system only uses the QR code on demand. Also, due to the fact that Bluetooth processes and matches local grouping information, and QR code requires a central server with limited privacy-preservation.

BeepTrace solution sits between them, by recording the information in the background, but only transmitting at a suitable time. For instance, while charging or docking the device, hence BeepTrace is not only privacy-preserving but also power preserving and battery friendly. Furthermore, BeepTrace brings a higher level of security to the user physical device as it overcomes the issue of Bluetooth wireless vulnerability and avoids the bureaucracy flaws in the health code system.

Meanwhile, since the Bluetooth only works locally, the coverage is also limited, whereas BeepTrace, using integrated services from the user and third party suppliers, the coverage can be boosted globally without much effort.

Besides, as opposed to centralized solutions or partially decentralized service, for example, Google and Apple need a central service to respond to APIs and geodata matching, which is risky as the access tracking is possible. BeepTrace demonstrates the incomparable benefits of security and privacy preservation as a completely decentralized service. While BeepTrace employs third-party servers for matching, it keeps the user privacy protected and preserved thanks to the passive listening mechanism to avoid triggering access tracking. 
In addition to the power and privacy concern, BeepTrace is a unique solution to handle the user's location with full life cycle care without giving up on privacy.

Next, we will numerically analyze the BeepTrace performance in terms of the storage at blockchain, computing complexity at the server, and data at the user.


\subsection{Blockchain performance requirement}
\begin{figure}
    \centering
    \includegraphics[scale = 0.60]{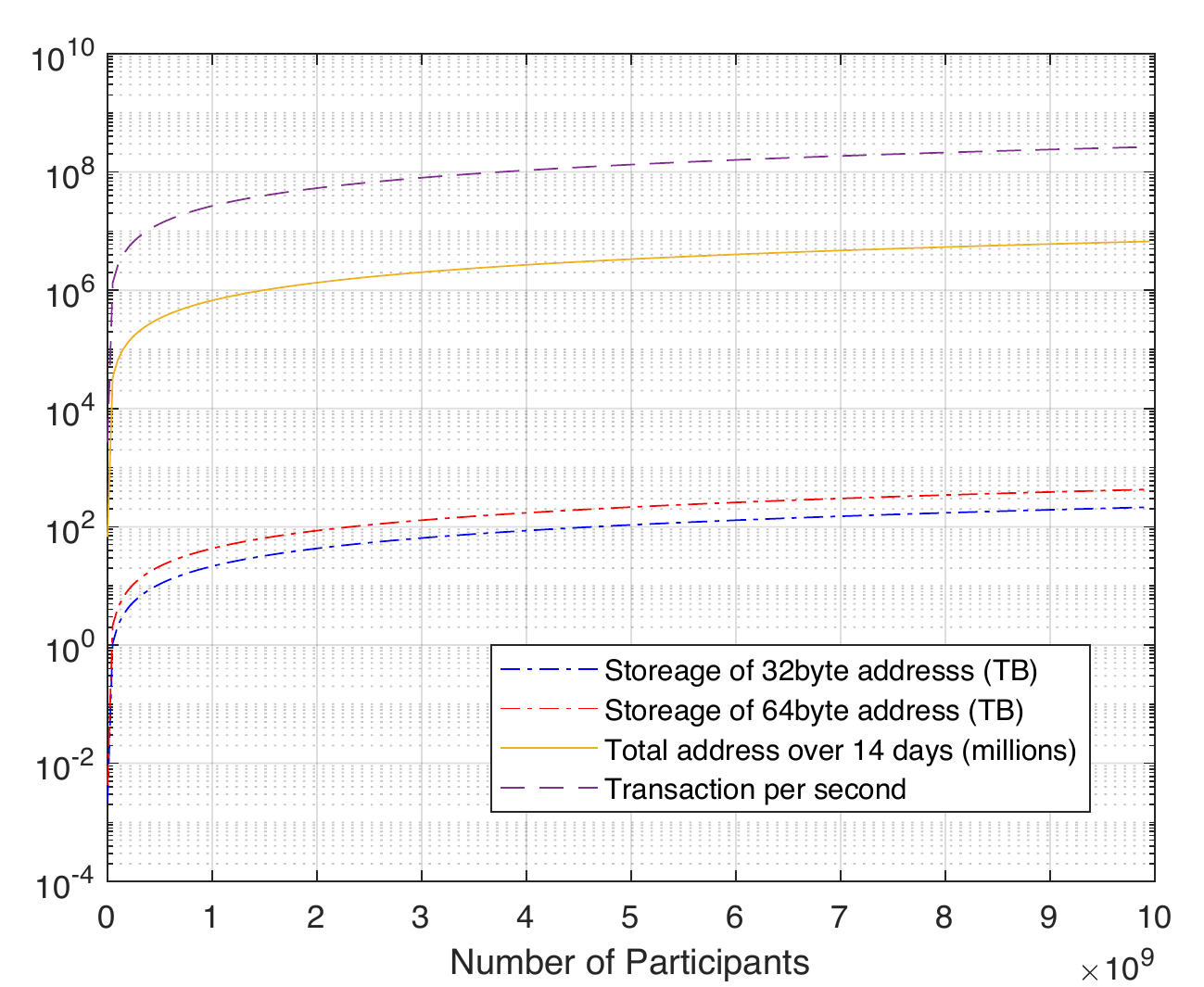}
    \caption{Storage and Capacity of Contact Tracing Blockchain address and transactions}
    \label{fig:storeage}
\end{figure}

Storing massive amounts of blockchain addresses is a bold challenge of contact tracing blockchain, due to the accumulating data uploaded by users. Therefore, a certain period of lifespan should be considered for the application of such a system. 
Thankfully, the contact tracing only requires a certain number of days of records (14 days only for COVID-19 according to WHO, which will be used as an example later), hence any data older than that number can be discarded. By estimating the number of participants and the size of each blockchain address, we plot in Fig. \ref{fig:storeage} the maximum allowed storage of 14 days with new addresses declared every 30 minutes from each user, against the number of participants. In addition, we plot the number of TPS against an increasing number of users.
The lines in blue and red compare the capacity between using 512 bit (64byte) and 256 bit (32byte) address, both of them end at approximately 200TB of data. 
It is worth pointing out that only the geodata solver requires such an amount of data for problem-solving, but for miners, the required blocks can be set to the newest dozens of blocks, which may just take a few megabytes.

In terms of notification blockchain, at the scale of 10,000 confirmed cases per day (see details in user side analysis), the total amount of storage required is far less (a few GB comparing with several hundreds of TB).

Transaction per second is a critical performance metric for the blockchain network. Using the assumption above, each user generates 1 address in 30 minutes, with $N$ users, the addresses uploaded to the system per second is calculated as $N\times \frac{1}{30\times 60}$. Such an amount of TPS is the core challenge for BeepTrace deployment on a large scale, hence mitigation should be considered to address the large TPS threshold, which is detailed in Section \ref{sec:challenges}.
\subsection{Geodata computing resource requirement}
Once the users start uploading their geodata, the server is involved with the job of geodata matching. It is a simple job of looking up the geodata coordinates and comparing it with all existing records. In this process, we define the workflow of address lookup and match as: 1. Read one of the confirmed patient's geodata; 2. Compute the distance between it and all records from the latest one and; 3. Make a transaction to the relevant address with risk level endorsement.

\begin{figure}
    \centering
    \includegraphics[scale = 0.65]{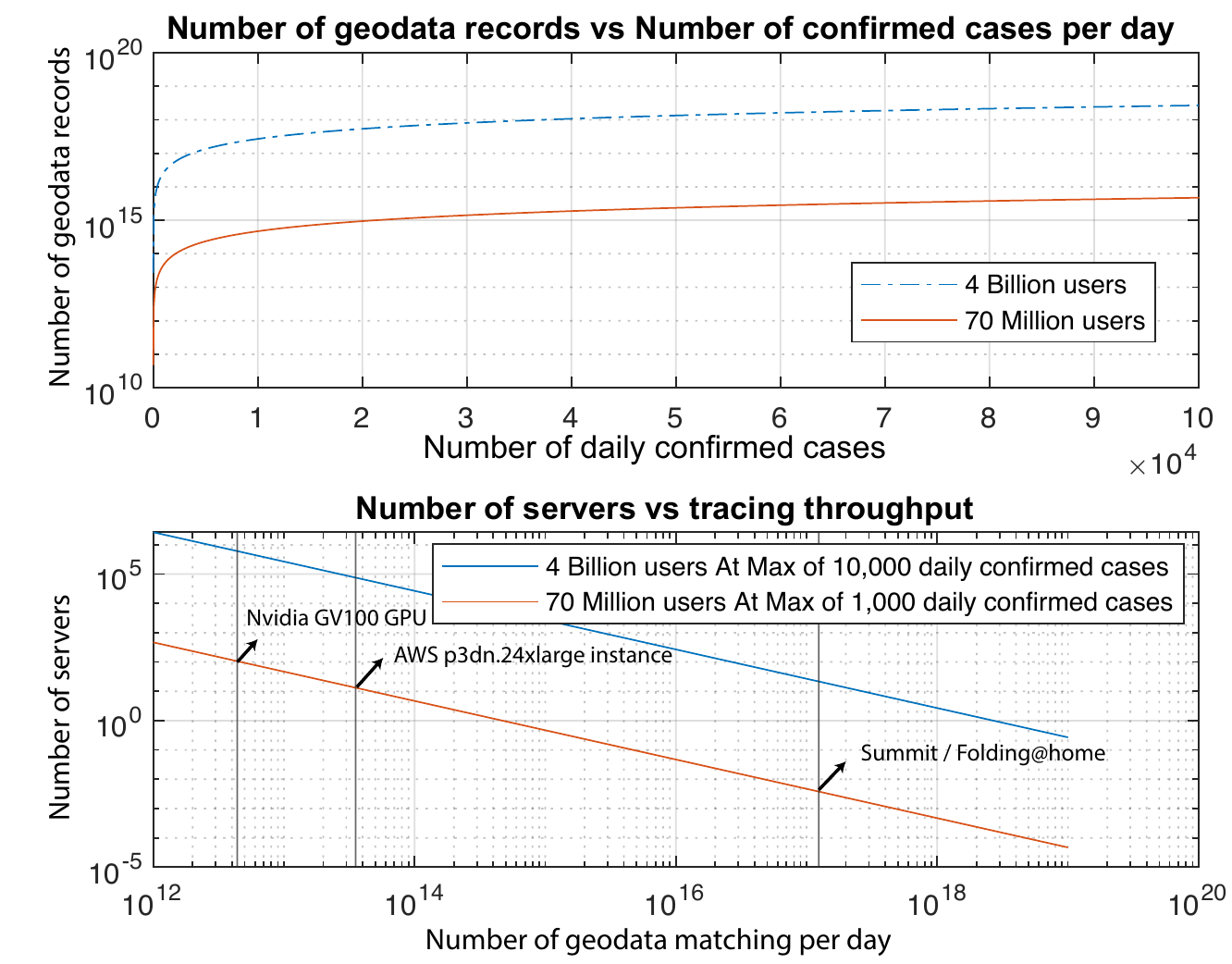}
    \caption{Computing Resource Requirement for Contact Tracing Geodata matching}
    \label{fig:computing}
\end{figure}

It is reasonable to assume that address lookup would take less than 0.1 ms a record (as a baseline, many modern processors are faster) on single CPU thread or CUDA core \cite{benchmarks}. Hence we can obtain an estimation of computing resource requirement against the number of users and daily confirmed cases. In Fig. \ref{fig:computing}(a), using the same setup from the previous simulation, i.e., 30 minutes an address regularly in 24 hours a day, we can see the number of records increases linearly with the number of daily confirmed cases, but exponentially with the overall participants' number. It shows that the system is linear against increasing confirmed cases but not very scalable if the network gets larger. 

The scalability is also simulated with the above parameters in Fig. \ref{fig:computing}(b), where the large tracing network needs more computing resources by comparing a system of 70 million and 4 billion population. It is evident that the cost of maintaining a network consisting of billion of users with Quint-scale message counts is not practical with current technology but completely manageable if the network scales down. In fact, for a medium-sized country with 70 million population (For example, UK, France), the cost of geodata solving is manageable with a handful of high-performance cluster servers, which cost far less than what is required for the larger network. A detailed breakdown of scalability challenges can be found in Section \ref{sec:challenges}.

\subsection{User-side resources requirements }
\begin{figure}
    \centering
    \includegraphics[scale = 0.6]{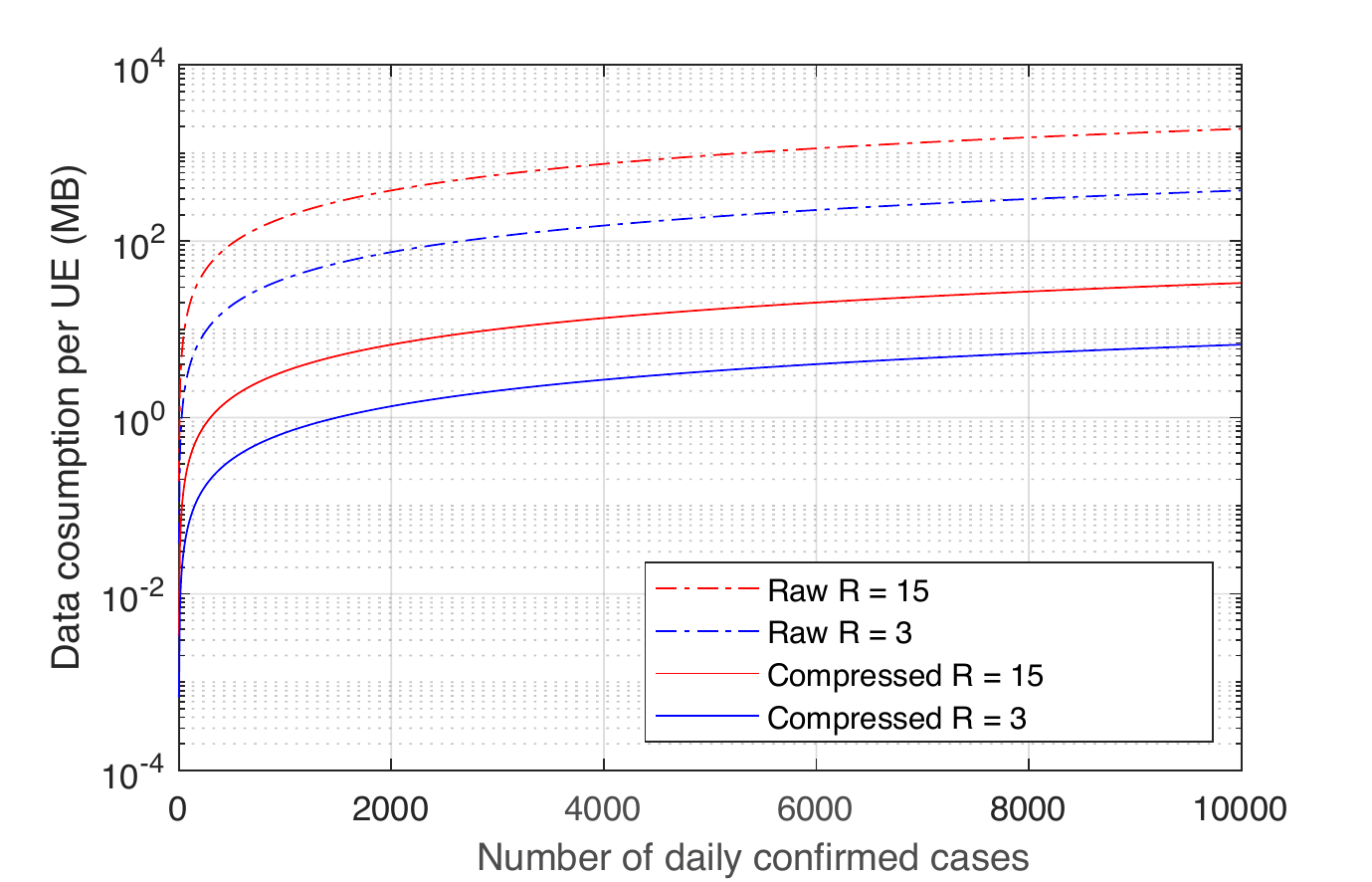}
    \caption{User data consumption}
    \label{fig:ue}
\end{figure}

Every user is considered as a thin-node of the blockchain tracing network, so the retrieving data and lookup records will take place on the local user and local user only, where the user privacy is preserved. It is going to consume the user's computing, storage, and network resources. In BeepTrace, the geodata solver marked addresses are announced on the notification blockchain, which is exclusive to the matching results. Assuming in each location it stayed, there will be $R=15$ blockchain addresses associated with the patient's geodata. Therefore, assume that the patient was traveling constantly and stayed in different places every 30 minutes in 14 hours (typical active time for an adult during a day). It means each patient will incur 420 records for the geodata solving and accumulating up to 5880 records for a 14 days interval. Note that, in the real world situation, the patient might interact with hundreds or even thousands of addresses in one time at one address, but only the most closed contact will lead to infection, hence the number of users being tagged by the geodata solver will be significantly less than the crowds the patient interacted. By adding all the records up, there is a challenge of processing and storing such a large volume of data, as illustrated in Fig. \ref{fig:ue}. The raw data recorded on the blockchain will be too heavy for the user side, therefore, we have to compress the associated blockchain addresses with MD5 checksum\cite{Rivest1992} or other fingerprints to reduce the size of files. Meanwhile, the solver side will need to employ a mechanism to remove duplicates and produce a single risk level endorsement for all the match geodata that associated with a single pseudonym. The number of records can be reduced to 210 and 16byte for each address's fingerprint, which is calculated based on a 14 hour active time per day in 14 days (14 x 15) since the daily records of 28 are summarized by the solver into one record, and pseudonym changes daily. By compressing the data using the above methods, we can see a dramatic drop in data consumption, 33.6MB for $R=15$, 6.7MB for $R=3$ per day, hence enabling a wider user range and minimizing the users' cost. More details regarding storage optimization can be found in Section \ref{sec:challenges}.
As for the uploading cost, the total amount of data upload to tracing blockchain throughout a day will be 28 addresses of 64byte, which is considerably less than download from notification blockchain.

\section{Challenges and Discussions}
\label{sec:challenges}
\subsection{Network throughput and scalability}
The major issues with our proposed contact tracing scheme are the massive traffic caused by a large amount of addresses declaration due to frequent (globe) geodata update, and the computing resources required for geodata matching. Meanwhile, we face a great challenge of blockchain processing throughput for single-chain operation. It is a great challenge running the desired hundreds of millions of transactions per second on any existing blockchain solution. Luckily, the needs of such high TPS is rare in the real world, for example, it is reasonable to assume that a user does not travel internationally often, therefore the needs of the user data are completely met in the domestic blockchain network. 
In addition, all parameters are selected at typical maximum values to see the peak requirement. For instance, it is not reasonable to assume all people (in all ages) are active 14 hours every day. We also encourage the use of multiple blockchains by regionally grouping the users via PKI and public keys management. By dividing users into smaller groups, the network capacity can be easily scaled up. Besides, the emerging high throughput ready blockchain can be introduced to the deployment of BeepTrace, for instance, DAG (Directed acyclic graph) in theory has no throughput limit thanks to its intentionally designed forking schemes \cite{Cao2019}. And when the technology is ready for high throughput performance, we can easily migrate two or more regional chains together and speed up the sharing of the information. 

The computing resources are limited from time to time, however, the geodata complexity can be dramatically reduced if the user's quantity on a single chain is below thresholds. In the case of international passengers, the country can employ the server to look up the data in both regional networks, hence reduces the needs of massive networks in all time. We have made a comparison of simulations based on an assumption of different size networks. 
For a medium-sized country with 70 million population, the required computing resource is as little as dozens of AWS EC2 \textit{p3dn.24xlarge} instances, however, for the large population bases, such as the combined population match of the top 7 most populated countries (a sum of 4 billion people), it takes tremendous computing resources equaling to 23 of Summit\cite{OakRidgeNationalLaboratory2018} (the fastest supercomputer in 2019) and hardly achievable using current technology, though it will be possible in the near future.

\subsection{Battery drainage and storage optimization}
All the recently proposed contract tracing programs have the challenge of battery drainage and storage optimization, which are not avoidable due to the requirements of active broadcast and recording of GPS coordinates.
However, our scheme can be more energy efficient by separating the recording and uploading in two steps. The user can store the recorded geodata on the local device and wait until it is plugged in and within the WiFi coverage. By sending the data only when the mobile device is being charged, our scheme becomes more battery-friendly. Delaying the information upstream can induce lower performance in the contact tracing network, but it is completely acceptable to be notified a few hours later rather than immediate response due to the nature of tracing lag. Also, geodata generation is paused if the user's locations remain the same, which also reduces the entries to the blockchain. From the solver side, if the duplicates of endorsement are made, the solver will only upload the unique address to the blockchain, which reduces the pressure on the user.
\subsection{Technology for elders and minors}\label{sec:elders}
Technology has certain advantages to the general public, however, elders and minors are often left out. The limits of technology reach to certain groups of people may become a major issue at rolling out digital contact tracing. But it is not completely impossible to include them. Wearable technology and wireless IoT\cite{Sun2019} can be used by the elders and minors to enable them for the contact tracing program. Under the scheme developed in the earlier section, the private keys were stored locally, but are transferable to guardians and carers. By transferring the private keys in a secured D2D channel\cite{D2D}, the parents and carers can take responsibility to keep their beloved under protection, without giving up on their privacy.

As discussed earlier, the risk level assessment is notified only via a passive broadcast, however, it is not limited to the passive-only situation. It is very likely that elders and minors will not be putting enough effort to receive the notifications, hence a trusted third party is needed in this case. By giving consent of privacy to some other users or third-party service providers, they can start sending push messages to the vulnerable once there is a risk. People naturally do give their privacy consent to the above parties, for example, care homes, online health companies, parents and adult children of elders. By combining these avenues, we believe no one should be left out in this crisis.
\subsection{Economical and social aspects} 
It is well-known that centralized systems are more efficient and economical than decentralized systems in most cases. Blockchain is a representative of distributed systems and deploying such a system in a nationwide manner may cost taxpayers more. However, from another side, the decentralized blockchain system is also well-recognized among citizens as a non-governmental solution that can preserve privacy in a much better way than a centralized system. Such a consensus can effectively minimize the resistance from human rights organizations and fear of citizens of infringing rights or other fundamental civil liberties. This will increase the uptake of the digital contact tracing among the citizens and is thus of paramount importance to winning the battle with COVID-19 as early as possible and to save billions each day.

From the blockchain mining perspective, attracting sufficient independent miners to contribute the blockchain construction is key to maintaining its nature of the distribution. In the most successful blockchains such as Bitcoin, the reward to the miners are from the transaction fees and/or creating a new block. In BeepTrace, it could be difficult to build such an ecosystem in a short time and there are no real transactions (thus no transaction fees) in such a system. As solutions, the reward can come from the government by paying the miners who created and maintained the blockchain, or in the case of sharing some existing blockchains, transaction fees can be claimed back from the government. Of course, conquering COVID-19 is the common mission of all mankind thus each user could be part of the miners to voluntarily support, legitimize, and monitor the blockchain network.

\section{Conclusions}
\label{sec:Conclusions}
A blockchain-enabled solution is proposed to solve the critical privacy-preserving issues in digital contact tracing for COVID-19 pandemic. The blockchains are enabled between the user/patient and the authorized solvers to desensitize the geodata from the user identity. Detailed procedures and functions of each entity are presented and compared with existing solutions to show the advantages. Challenges are also discussed from blockchain performance, solvers complexity, user's battery and storage, economic and social aspects, respectively. Our numerical results show that the proposed BeepTrace is all around winner from security, privacy, battery, coverage perspectives. This solution provides an in time framework for governments, authorities, companies, and research institutes over the world to develop a trusted platform for tracing information sharing, to win the fight with the COVID-19 pandemic.  

\bibliographystyle{IEEEtran}
\bibliography{IEEEabrv,reference}

\end{document}